\documentclass[12pt]{article}

\usepackage[a4paper,left=30mm,right=20mm,top=20mm,bottom=20mm]{geometry}
\usepackage{graphics}
\usepackage{amsmath}
\usepackage{amssymb}
\usepackage{epsfig}
\usepackage{cite}
\usepackage{xcolor}
\usepackage[unicode]{hyperref}
\newcommand{\bea}{\begin{eqnarray}}
\newcommand{\eea}{\end{eqnarray}}
\newcommand{\be}{\begin{equation}}
\newcommand{\ee}{\end{equation}}
\newcommand{\MSbar}{\overline{\rm MS}}

\title{TMD gluon density in nuclei versus experimental data on heavy flavor production at LHC}

\author{ A.V.~Lipatov$^{1,2}$, M.A.~Malyshev$^{2}$, A.V.~Kotikov$^{1}$, X.~Chen$^{3,4}$}

\begin{document}

\maketitle

\begin{center}

{\it $^{1}$Joint Institute for Nuclear Research, 141980, Dubna, Moscow region, Russia}\\
{\it $^{2}$Skobeltsyn Institute of Nuclear Physics, Lomonosov Moscow State University, 119991, Moscow, Russia}\\
{\it $^{3}$Institute of Modern Physics, Chinese Academy of Sciences, Lanzhou 730000, China}\\
{\it $^{4}$School of Nuclear Science and Technology, University of Chinese Academy of Sciences, Beijing 100049, China}\\

\end{center}

\vspace{0.5cm}

\begin{center}

{\bf Abstract }
       
\end{center}

\indent

Analytical expressions for the Transverse Momentum Dependent (TMD, or unintegrated) gluon and sea quark
densities in nuclei are derived at leading order of QCD running coupling.
The calculations are performed in the framework of the rescaling model and
Kimber-Martin-Ryskin (KMR) prescription, where the Bessel-inspired behavior of parton densities at 
small Bjorken $x$ values, obtained in the case of flat initial conditions in the
double scaling QCD approximation, is applied. 
The derived expressions are used to evaluate the 
inclusive heavy flavor production in proton-lead collisions at the LHC. %and corresponding nuclear modification factor.
We find a good agreement of our results with 
latest experimental data collected by the CMS and ALICE Collaborations at $\sqrt s = 5.02$~GeV.

\vspace{1.0cm}


\newpage

\section{Introduction} \noindent

It is well known that the
usual concept of QCD factorization for proton-proton ($pp$) interactions assumes that the corresponding inclusive production
cross section is calculated as a convolution of short-distance partonic
cross sections and parton (quark or gluon) distribution functions in a proton (PDFs).
Suggesting that there is no hot medium formed in proton-nucleus ($pA$) collisions,
this concept is often extrapolated to $pA$ interactions by replacing usual PDFs with nuclear 
PDFs (nPDFs) while keeping hard scattering cross sections the same (see, for example,\cite{EPPS22, nNNPDF3, nCTEQ15, nCTEQ15upd}).
However, some additional phenomena, so called cold nuclear matter effects\footnote{These are, for example, soft gluon interactions between 
hadrons in the initial and final states.}, can affect this picture (see\cite{nFactorization-1, nFactorization-2, nFactorization-3, nFactorization-4}
and references therein).
Detailed knowledge of nPDFs, in particular, gluon distribution in nuclei, is necessary for theoretical description
of $pA$ processes studied at modern (LHC, RHIC) and future colliders (FCC-he, EiC, EicC, NICA). Moreover, it is important
in discriminating the initial nuclear effects from the subsequent hot medium
effects essential for more complex nucleus-nucleus ($AA$) collisions,
where nuclear matter can reach extremely high energy densities and temperatures, transforming
into quark-gluon plasma (see also review\cite{Nuclear-Review} for more information).
In this sence, production of charm and beauty flavors in $pA$ interactions is of particular interest, 
because these processes directly probe the gluon distributions in the colliding particles.

Experimental investigation of deep inelastic scattering of leptons on nuclei 
performed by the European Muon Collaboration reveals
the appearance of a significant nuclear effect\cite{EMC}, which excludes the naive idea of the nucleus as
a system of quasi-free nucleons (see also\cite{EMC-Review-1, EMC-Review-2} for review).
From theoretical point of view, there are two main scenarios to determine the nPDFs.
In the first of them, nPDFs at some initial (or starting) scale $\mu_0^2$ are extracted from global fits to nuclear 
data using empirical parametrization of their shape and normalizations (see, for example,\cite{EPPS22, nNNPDF3, nCTEQ15, nCTEQ15upd} and references therein). Then, 
numerical solution of Dokshitzer-Gribov-Lipatov-Altarelli-Parisi (DGLAP) equations\cite{DGLAP} 
is employed to describe the corresponding QCD evolution (dependence on the scale $\mu^2$).
An alternative approach is based on some nPDF models (see\cite{Kulagin} for more information). 
Here we follow the rescaling model\cite{RescalingModel} based on the assumption that
the effective size of gluon and quark confinement in the nucleus is greater than in the free nucleon\cite{RescalingModelAssumptions}.
Within the perturbative QCD, this confinement rescaling predicts that PDFs and nPDFs 
can be connected by scaling the argument $\mu^2$ (see also review\cite{RescalingModel-Review}). 
So that, the rescaling model demonstrates the features inherent in both approaches: there are 
certain relationships between conventional and nuclear PDFs that arise as a result of shifting 
the values of the scale $\mu^2$ and, at the same time, both densities obey DGLAP equations.
Initially, the rescaling model was proposed for the domain of valence quarks dominance, $0.2 \leq x \leq 0.8$, 
where $x$ is the Bjorken variable. Later
it was extended to small $x$\cite{RescalingModel-Extension-1, RescalingModel-Extension-2} (see also a short review\cite{RescalingModel-Extension-3}),
where certain shadowing and antishadowing effects were found for gluon and sea quark distributions\footnote{Recently the 
rescaling model was applied to the linealy polarized gluon density in nuclei\cite{RescalingModel-LinearlyPolarizedGluons}, see also\cite{LinearlyPolarizedGluons}.}.

The main goal of our study is to derive analytical expressions (at leading order in the QCD coupling $\alpha_s$) for the 
Transverse Momentum Dependent (TMD, or unintegrated) gluon and sea quark densities in nuclei (nTMDs)
using the rescaling model\cite{RescalingModel} and popular Kimber-Martin-Ryskin (KMR) formalism\cite{KMR-LO} (see also\cite{KMR-NLO}). 
These quantities encode nonperturbative information on hadron structure, including transverse momentum and polarization degrees of freedom.
Currently they are widely used in a number of 
applications to topical issues in high energy physics phenomenology,
especially for multi-scale or non-inclusive collider observables (see, for example, review\cite{TMD-Review} for more information).
Next, the calculated TMD gluon density
will be applied
to investigate the heavy flavor (charm and beauty) production in proton-lead collisions at $\sqrt s_{\rm NN} = 5.02$~TeV.
Such processes are known to be sensitive to the gluon content of the nucleus and
provide us with possibility to reconstruct the full map of the latter\cite{TMD-map}.
Of course, they are great of importance to first test of the derived expressions.
We will use the $k_T$-factorization\cite{kt-factorization}, or high-energy factorization\cite{HighEnergyFactorization} approach,
which turns to be a convenient alternative to explicit higher-order pQCD calculations.
In fact, a large piece of NLO~+~NNLO~+~... corrections important at high energies
can be effectively taken into account in the form of the TMD gluon density\cite{TMD-Review}.
Our predictions are compared with latest experimental data taken by the
CMS\cite{CMS-B,CMS-bjets,CMS-cjets} and ALICE\cite{ALICE-D} Collaborations at the LHC.
The consideration below extends and continues the line of our previous studies\cite{RescalingModel-LinearlyPolarizedGluons, LinearlyPolarizedGluons, TMD-DAS-1, TMD-DAS-2}.

The outline of our paper is following. In Section~2 we briefly describe our theoretical
framework and basic steps of our calculation. In Section~3 we present the numerical
results and discussion. Section~4 sums up our conclusions.

\section{The model} \noindent

This section provides a 
short description of the calculation steps for the TMD gluon 
and sea quark densities in a proton and nuclei
and a brief review of the $k_T$-factorization formulas
for heavy flavor production. 

\subsection{Conventional (collinear) PDFs in a proton} \noindent

As it was argued\cite{DAS-Theory-1}, the HERA small-$x$ data can be well interpreted in terms
of so-called double asymptotic scaling approximation, which is related to the asymptotic behaviour of DGLAP evolution (see also\cite{DAS-Theory-2, DAS-Theory-3, DAS-Theory-4}).
In this approximation, flat initial conditions for conventional gluon and sea quark densities in a proton
$f_a(x, \mu^2)$ at some scale $Q_0^2$ could be used: $f_a(x, Q_0^2) = A_a$, where $a = q$ or $g$\cite{PDFs-DAS-1, DAS-Theory-2, DAS-Theory-3}. 
At leading order (LO) of perturbative QCD, small-$x$ expressions for $f_a(x, \mu^2)$ read\footnote{The both leading and next-to-leading formulas and their derivation 
can be found\cite{PDFs-DAS-1, DAS-Theory-2}.}:
\begin{gather}
f_a(x,\mu^2) = f^{+}_a(x,\mu^2) + f^{-}_a(x,\mu^2), \nonumber \\
f^{+}_g(x,\mu^2) = \biggl(A_g + {4\over 9} A_q \biggl) \overline{I}_0(\sigma) e^{-\overline d_{+} s} + O(\rho), \quad f^{-}_g(x,\mu^2) = - {4\over 9} A_q e^{- d_{-} s} + O(x), \nonumber \\
f^{+}_q(x,\mu^2) = \frac{N_f}{3} \biggl(A_g + {4\over 9} A_q \biggl) \tilde{I}_1(\sigma)  e^{-\overline d_{+} s} + O(\rho), \quad f^{-}_q(x,\mu^2) =  A_q e^{-d_{-} s} + O(x),
\label{8.02}
\end{gather}
\noindent
where %$C_A=N_c$, $C_F=(N_c^2-1)/(2N_c)$ for the color $SU(N_c)$ group, ($\nu=0,1$)
\begin{gather}
s=\ln \left( \frac{\alpha_s(Q^2_0)}{\alpha_s(\mu^2)} \right), \quad \sigma = 2\sqrt{\left|\hat{d}_+ s\right| \ln \left( \frac{1}{x} \right)}, \quad \rho=\frac{\sigma}{2\ln(1/x)}, \nonumber \\
\hat{d}_+ = - \frac{12}{\beta_0}, \quad \overline d_{+} = 1 + \frac{20 N_f}{27\beta_0}, \quad d_{-} = \frac{16 N_f}{27\beta_0},
\label{intro:1a}
\end{gather}
\noindent
and $\overline{I}_{\nu}(\sigma)$ and
$\tilde{I}_{\nu}(\sigma)$ 
are combinations of modified Bessel functions (at $s\geq 0$, i.e. $\mu^2 \geq Q^2_0$) and usual
Bessel functions (at $s< 0$, i.e. $\mu^2 < Q^2_0$):
\begin{gather}
\tilde{I}_{\nu}(\sigma) =
\left\{
\begin{array}{ll}
\rho^{\nu} I_{\nu}(\sigma) , & \mbox{ if } s \geq 0; \\
(-{\rho})^{\nu} J_{\nu}({\sigma}) , & \mbox{ if } s < 0,
\end{array}
\right. \quad
\overline{I}_{\nu}(\sigma) =
\left\{
\begin{array}{ll}
\rho^{-\nu} I_{\nu}(\sigma) , & \mbox{ if } s \geq 0; \\
{\rho}^{-\nu} J_{\nu}({\sigma}) , & \mbox{ if } s < 0.
\end{array}
\right.
\label{4}
\end{gather}
\noindent
Here $N_f = 4$ is the number of active (massless) quark flavors and $\beta_0 = 11 -2 N_f/3$ is the first coefficient of the QCD
$\beta$-function in the $\MSbar$-scheme.
The parameters $A_a$ and $Q_0^2$ have been extracted\cite{PDFs-DAS-1} from a fit to HERA data on the proton structure function $F_2(x,Q^2)$
for $\alpha_s(M_Z^2) = 0.1168$ (see below).

\subsection{Rescaling model and nuclear PDFs} \noindent

In the rescaling model\cite{RescalingModel}, the structure function $F_2^A(x, Q^2)$
and, consequently, the valence part $f_{V}^A(x,\mu^2)$ of the quark density in a nucleus $A$ are modified 
at intermediate and large $x$ values, $0.2 \leq x \leq 0.8$, as follows
\begin{gather}
  f_{V}^A(x,\mu^2) = f_{V}(x,\mu^2_{A,V}),
  \label{va.1}
\end{gather}
\noindent
where the new scale $\mu^2_{A,V}$ is related to $\mu^2$ by\cite{RescalingModel-Extension-1}
\begin{gather}
s^A_V \equiv \ln {\ln \mu^2_{A,V}/\Lambda_{\rm QCD}^2 \over \ln Q^2_{0}/\Lambda_{\rm QCD}^2} = s + \ln( 1 + \delta^A_V) \approx s +\delta^A_V.
\label{sA}
\end{gather}
\noindent
So that, kernel modification of the main variable $s$ defined in~(\ref{intro:1a}) depends on the
$\mu^2$-independent parameter $\delta^A_V$ having small values\cite{RescalingModel-Extension-1}.

Next, since rise of sea quark and gluon densities increases with increasing values of $\mu^2$,
the small-$x$ PDF asymptotics~(\ref{8.02}) 
were applied\cite{RescalingModel-Extension-1} to the small $x$ region of the EMC effect.
In fact, in the case of nuclei, the evolution scale is less than $\mu^2$, that 
can directly reproduce the shadowing effect observed in global fits.
So, one can assume that
\begin{gather}
  f^{A}_a(x,\mu^2) = f_a^{A,+}(x,\mu^2) + f_a^{A,-}(x,\mu^2), \quad f^{A,\pm}_a(x,\mu^2) = f^{\pm}_a(x,\mu^2_{A,\pm}),
\label{AD1}
\end{gather}
\noindent
where $f_a^\pm(x, \mu^2)$ are given by~(\ref{8.02}).
Thus, there are two free parameters $\mu^2_{A,\pm}$ which should be determined from the analysis of
experimental data for the EMC effect at low $x$.
Corresponding values of $s^{A}_{\pm} $ turned out to be
\begin{gather}
s^{A}_{\pm} \equiv \ln {\ln \mu^2_{A,\pm}/\Lambda_{\rm QCD}^2 \over \ln Q^2_{0}/\Lambda_{\rm QCD}^2} = s +\ln(1+\delta^{A}_{\pm}),
\label{AD2}
\end{gather}
\noindent
where $\delta^{A}_{\pm}$ can be presented as\cite{RescalingModel-Extension-1}
\begin{gather}
  -\delta^{A}_{\pm}=c^{1}_{\pm}+c^{2}_{\pm} A^{1/3}
\label{dpm}
\end{gather}
\noindent
and 
\begin{gather}
c^{1}_{+}=-0.055\pm 0.015, \quad c^{2}_{+}=0.068\pm 0.006, \nonumber \\
c^{1}_{-}=0.071\pm 0.101, \quad c^{2}_{-}=0.128\pm 0.039.
\label{cpm}
\end{gather}
\noindent
In particular, for $^{208}$Pb we have $\delta^{\rm Pb}_{+}=-0.34$ and $\delta^{\rm Pb}_{-}=-0.78$.

\subsection{Kimber-Martin-Ryskin approach} \noindent

The KMR approach is a
formalism to construct the TMD gluon and quark distributions from conventional (collinear) PDFs. The key assumption here is that
the transverse momentum dependence of the parton densities  
enters only at the last of QCD evolution (namely, DGLAP). The KMR procedure is believed to take into account 
effectively the main part of next-to-leading logarithmic (NLL) terms $\alpha_s^n \ln^{n-1} \mu^2/\Lambda_{\rm QCD}^2$
compared to the leading logarithmic approximation (LLA), where terms proportional to $\alpha_s^2 \ln^n \mu^2/\Lambda_{\rm QCD}^2$
are taken into account.

In the integral formulation of KMR approach,
the TMD gluon and quark densities at the leading order\footnote{The next-to-leading formulas can be found\cite{KMR-NLO}.} of $\alpha_s$ can be written as\cite{KMR-LO}
\be
f_a(x,{\mathbf k}^2_T,\mu^2) = {\alpha_s({\mathbf k}_T^2) \over 2\pi {\mathbf k}_T^2} T_a(\mu^2,{\mathbf k}^2_T) \sum_{a'} \int\limits^{1 - \Delta({\mathbf k}^2_T)}_x \frac{dz}{z} P_{aa'}^{(0)}(z)  D_a\left(\frac{x}{z},{\mathbf k}_T^2 \right),
\label{Def2}
\ee
\noindent
where $D_a(x,\mu^2) = f_a(x, \mu^2)/x$ are the conventional PDFs in a proton, 
$P_{a a'}^{(0)}(z)$ are the usual unregulated leading order DGLAP splitting functions and $a, a' = q$ or $g$. 
The Sudakov form factors $T_a(\mu^2,{\mathbf k}^2_T)$ enable one to include 
logarithmic loop corrections and have the following form
\be
T_a(\mu^2,{\mathbf k}_T^2)= \exp \left\{ - \int\limits^{\mu^2}_{{\mathbf k}_T^2} dp^2 \, { \alpha_s(p^2) \over 2\pi p^2} \sum_{a'} \int\limits^{1 - \Delta(p^2)}_0 \, dz \, z P_{a'a}^{(0)}(z) \right\}\, .
\label{Ta}
\ee
\noindent
The cut-off parameter $\Delta({\mathbf k}^2_T) = |{\mathbf k}_T|/(\mu + |{\mathbf k}_T|)$ imply the angular-ordering constraint\footnote{Another choice,
$\Delta({\mathbf k}^2_T) = |{\mathbf k}_T|/\mu$, which corresponds to the strong ordering condition, is also used in applications.}
specifically to the last evolution step to regulate soft gluon singularities.
Following\cite{PDFs-DAS-1}, everywhere below we use the phenomenological infrared modification of QCD coupling 
which effectively increases its argument at small scales,
namely, $\alpha_s(\mu^2) \to \alpha_s(\mu^2 + m^2_{\rho})$, where $m_\rho$ is the $\rho$ meson mass ('freezing' treatment), see\cite{frozen} and discussions therein.

\subsection{Analytical expressions for TMDs and nTMDs} \noindent

Using conventional PDFs given by~(\ref{8.02}) as an input for KMR procedure,
one can derive the analytical expressions for TMD gluon and quark densities in a proton.
After some algebra we have\cite{TMD-DAS-1}
\begin{gather}
f_a(x,{\mathbf k}^2_T,\mu^2) = {c_a \alpha_s({\mathbf k}_T^2) \over \pi {\mathbf k}_T^2 } T_a(\mu^2,{\mathbf k}_T^2) \times \nonumber \\
\times \left[ D_a(\Delta) f_a\left(\frac{x}{1 - \Delta},{\mathbf k}_T^2\right) + D_a^+ f_a^+\left(\frac{x}{1 - \Delta},{\mathbf k}_T^2\right) + D_a^{-} f_a^-\left({x\over 1 - \Delta}, {\mathbf k}_T^2\right) \right],
\label{uPDF2.1}
\end{gather}
\noindent
where 
\begin{gather}
D_q(\Delta)= \ln\left(\frac{1}{\Delta}\right) - {(1 - \Delta)(3 - \Delta) \over 4}, \quad 
D_g(\Delta)= \ln\left(\frac{1}{\Delta}\right) - {(1 - \Delta)(13 - 5 \Delta + 4 \Delta^2)\over 12}, \nonumber \\
D_q^-(\Delta) = - N_f \frac{(1 - \Delta)}{18} \left(2 - \Delta + 2 \Delta^2\right), \quad D_g^-(\Delta)=0, \nonumber \\
D_g^+= \frac{1}{\overline{\rho}_g} -(1 - \Delta) + \frac{(1 - \Delta)^2}{4} + \frac{4}{81} N_f, \nonumber \\
D_q^+=  {9 \over 8} (1 - \Delta) \left[ {2 \Delta^2 - \Delta + 2 \over {\rho}_a} - {4 \Delta^2 + \Delta + 13 \over 6} \right],
\label{uPDF2.3}
\end{gather}
\noindent
and 
\begin{gather}
\frac{1}{\rho_g} = {{ I_1(\sigma) \over \rho I_0(\sigma) } \vline}_{\,x \to x/(1 - \Delta)}, \quad
\frac{1}{\rho_q} = {{ I_0(\sigma) \over \rho I_1(\sigma) } \vline}_{\,x \to x/(1 - \Delta)}.
\label{rho.a}
\end{gather}
\noindent
Here 
$c_g = C_A = N_c$, $c_q = C_F = (N_c^2 - 1)/(2 N_c)$,
$\Delta \equiv \Delta({\mathbf k}_T^2)$ corresponds to the angular ordering constraint and 
'frozen' treatment of QCD coupling is implied.

Following\cite{TMD-DAS-1, TMD-DAS-2}, the analytic expression for TMD parton densities~(\ref{uPDF2.1}) % in a proton 
has to be modified at large $x$ in the form:
\begin{gather}
  f_a(x,{\mathbf k}^2_T,\mu^2) \to f_a(x,{\mathbf k}^2_T,\mu^2) \left(1 - {x \over 1 - \Delta} \right)^{\beta_a(s)}, \quad \beta_a(s) = \beta(0) + {4 c_a s\over \beta_0},
  \label{mod}
\end{gather}
\noindent
that is in agreement with similar modifications of conventional PDFs (see\cite{RescalingModel-Extension-1, RescalingModel-Extension-2} and references therein).
The value of $\beta_a(0)$ can be estimated from the quark counting rules
or extracted from the data.
Below we set $\beta_g(0) = 3.03$, which was derived\cite{TMD-DAS-1} from the best description of LHC data
on inclusive $b$-jet production in $pp$ collisions\footnote{The TMD gluon density in a proton given by~(\ref{uPDF2.1}) --- (\ref{mod}) 
is available in the popular \textsc{tmdlib} package\cite{TMDLib2} as KLSZ'2020 set.}.

\begin{figure}
\begin{center}
\includegraphics[width=7.9cm]{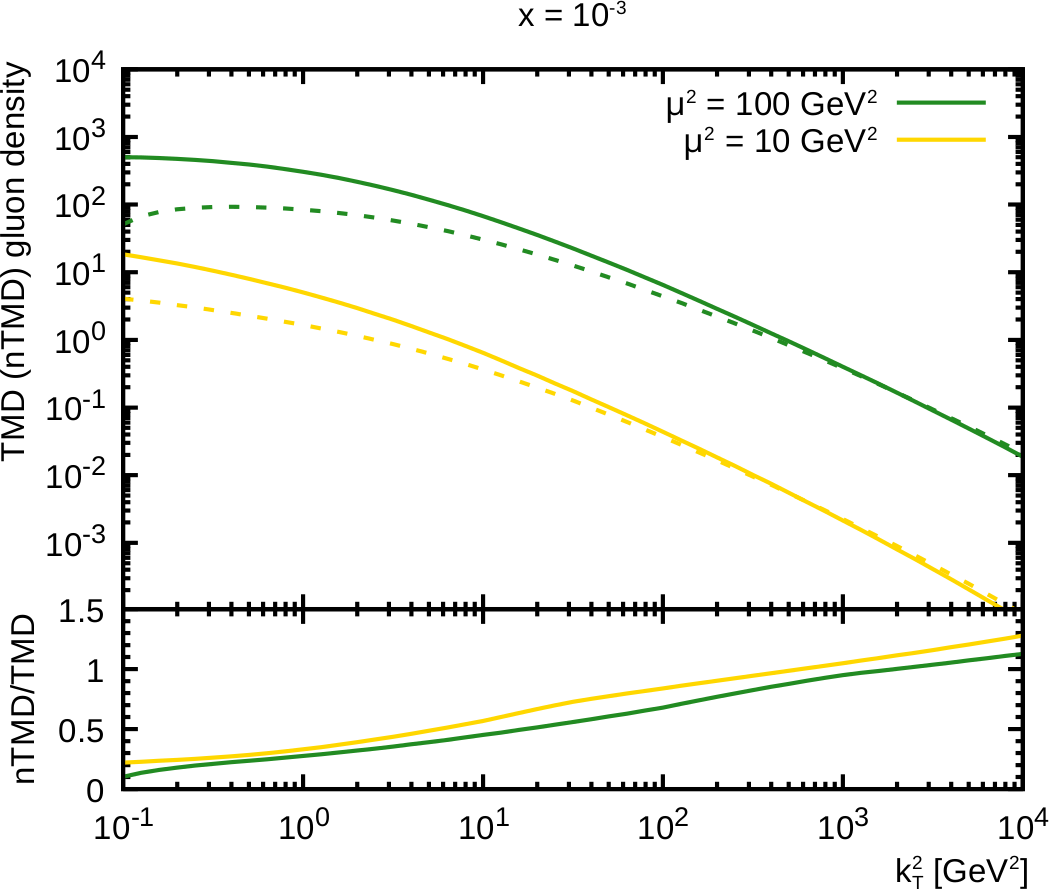}
\includegraphics[width=7.9cm]{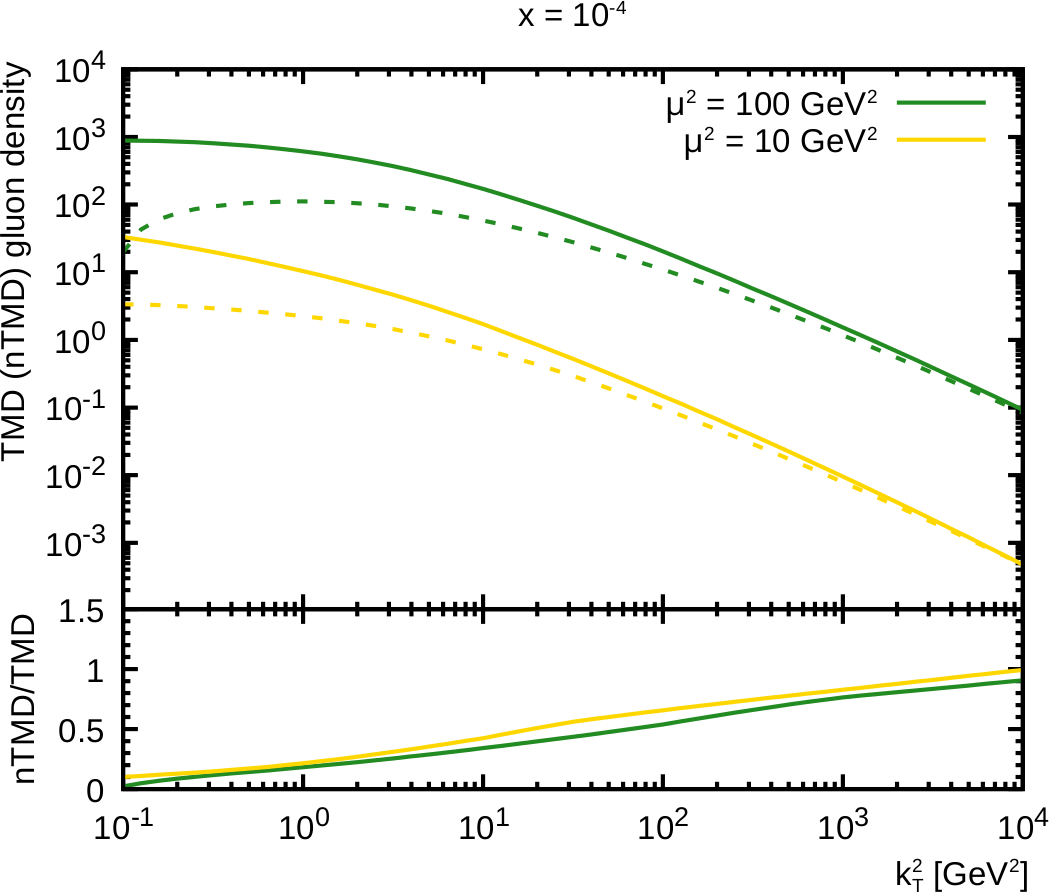}
\caption{The TMD gluon densities in a proton (solid curves) and nuclei (dashed curves) 
calculated as a function of ${\mathbf k}_T^2$ for different values of $x$ and $\mu^2$.
Note that results for $\mu^2 = 100$~GeV$^2$ are multiplied by a factor of $100$. The ratios
$f^{A}_g(x,{\mathbf k}_T^2,\mu^2)/f_g(x,{\mathbf k}_T^2,\mu^2)$ are presented also.}
\label{fig-tmds}
\end{center}
\end{figure}

Performing calculations in a similar way,
for TMD gluon and sea quark densities in nuclei %within the rescaling model 
we have
\begin{gather}
f^{A}_a(x,{\mathbf k}_T^2,\mu^2) = {c_a \over \pi {\mathbf k}_T^2 } \biggl[ \alpha_s({\mathbf k}_{A,+}^2) \left(D_a(\Delta)+D_a^+) T_a(\mu^2,{\mathbf k}_{A,+}^2\right) f^{A,+}_{a}\left(\frac{x}{1 - \Delta},{\mathbf k}_{A,+}^2\right) + \nonumber  \\
+ \alpha_s({\mathbf k}_{A,-}^2) \left(D_a(\Delta)+D_a^-\right) T_a(\mu^2,{\mathbf k}_{A,-}^2) f^{A,-}_{a}\left(\frac{x}{1 - \Delta},{\mathbf k}^2_{A,-}\right)\biggr],
\label{nuPDF2.1}
\end{gather}
\noindent
where $s \to s^A_{\pm}({\mathbf k}^2_T)$ and ${\mathbf k}^2_{A,\pm}$ could be easily derived from
\begin{gather}
  s^{A}_{\pm}({\mathbf k}_T^2) \equiv \ln { \ln {\mathbf k}^2_{A,\pm}/\Lambda^2_{\rm QCD} \over \ln Q^2_{0}/\Lambda^2_{\rm QCD}}
  = s +\ln(1+\delta^{A}_{\pm}) = \ln { \ln {\mathbf k}^2_{T}/\Lambda^2_{\rm QCD} \over \ln Q^2_{0}/\Lambda^2_{\rm QCD}} + \ln(1+\delta^{A}_{\pm}).
\label{AD2k}
\end{gather}
\noindent
The TMD gluon densities in a proton and nuclei calculated according to~(\ref{uPDF2.1}) and~(\ref{nuPDF2.1}) with the 'frozen' scenario for 
strong coupling and appropriate treatment of $\beta_g(0)$ are shown in Fig.~\ref{fig-tmds} as a function of gluon transverse momentum 
${\mathbf k}_T^2$ for different values of $x$ and hard scale $\mu^2$. 

\subsection{Cross section of heavy flavor production} \noindent

In the $k_T$-factorization approach, the heavy flavor production in $pp$ or $p\bar p$ collisions 
is dominated by the direct leading-order off-shell gluon-gluon fusion subprocess
\begin{gather}
  g^*(k_1) + g^*(k_2) \to Q(p_1) + \bar Q(p_2),
\label{subprocess}
\end{gather}
\noindent
where $Q = c$ or $b$ and four-momenta of corresponding particles
are given in the parentheses. 
The contribution from the quark induced subprocesses is of almost no importance
due to comparatively low quark densities. 
Corresponding cross section is calculated as a convolution of the off-shell (dependent on non-zero virtualities of the incoming gluons)
partonic cross section and TMD gluon distribution in a proton\cite{kt-factorization, HighEnergyFactorization}.
As it was already mentioned above, here we extend the consideration into $pA$ collisions by employing the master factorization formula:
\begin{gather}
  \sigma=\int\frac{dx_1}{x_1}\frac{dx_2}{x_2}\frac{d\phi_1}{2\pi}\frac{d\phi_2}{2\pi}d\mathbf k_{T1}^2\,d\mathbf k_{T2}^2f_g(x_1,\mathbf k_{T1}^2,\mu^2)f_g^A(x_2,\mathbf k_{T2}^2,\mu^2)d\hat\sigma^*(g^*g^*\to Q\bar Q),
  \label{HF}
\end{gather}
\noindent
where the initial off-shell gluons have fractions $x_1$ and $x_2$ of the parent proton and nucleus 
longitudinal momenta and azimuthal angles $\phi_1$ and $\phi_2$.
As usual, the off-shell partonic cross section reads
\begin{gather}
  d\hat \sigma^*(g^*g^* \to Q\bar Q) = {(2\pi)^4 \over 2 x_1 x_2 s} |\bar {\cal A}^2(g^*g^* \to Q\bar Q)| \times \nonumber \\ 
  \times {d^3 p_1 \over (2\pi)^2 2p_1^0} {d^3 p_2 \over (2\pi)^2 2p_2^0} \delta^{(4)}(k_1 + k_2 - p_1 - p_2).
\end{gather}
\noindent
The analytic expression for the off-shell gluon-gluon fusion amplitude
$|\bar {\cal A}^2(g^*g^* \to Q\bar Q)|$
is known for quite a long time (see, for instance,\cite{kt-factorization, HighEnergyFactorization, HF-ZLS}). 
Below we use it with the derived formulas~(\ref{uPDF2.1}) and (\ref{nuPDF2.1})
for TMD gluon densities in a proton and nuclei, respectively.
In all other respect our calculation is generally identical to that performed previously\cite{HF-JKLZ}.

\section{Numerical results and discussion} \noindent

We are now in a position to present our numerical results. First we describe our input
and the kinematic conditions. After we fixed the TMD gluon densities in a proton and nuclei, 
the cross section~(\ref{HF}) depends on the renormalization and factorization scales, $\mu_R$ and $\mu_F$. 
We take them to be equal to transverse mass of the leading produced heavy quark, $\mu_R^2 = \mu_F^2 = \xi^2(m_Q^2+\mathbf p_T^2)$, where the $\xi$ parameter 
is altered from $1/2$ to $2$ around its default value of $1$ to estimate theoretical uncertainties of our calculations. 
The quark masses are taken as $m_c=1.4$~GeV and $m_b=4.75$~GeV. The calculations were made with the one-loop formula 
for QCD coupling $\alpha_s$ with $\Lambda_\text{QCD}=143$~MeV, $Q_0^2 = 0.43$~GeV$^2$ and
%the number of active quark flavors of $N_f=4$. 
%According to fit\cite{?}, we set
$A_g = 0.77$, $A_q = 0.99$\cite{PDFs-DAS-1,PDFsSumRules} for 'frozen' treatment of $\alpha_s$.
In case of heavy jet production, 
we fully associate the 
produced heavy quark with the final jet. %in this work. %, such a simplification seems to be quite reasonable (see examples in~\cite{HF-JKLZ}). 

%The mesons are obtained from the heavy quarks in our calculations with standard Peterson fragmentation function with parameters $\epsilon_c=0.06$ and $\epsilon_b=0.006$ and the
%branching ratios are taken as $Br(c\to D)=0.559$ and $Br(b\to B)=0.398$.

\begin{figure}
\begin{center}
\includegraphics[width=7.8cm]{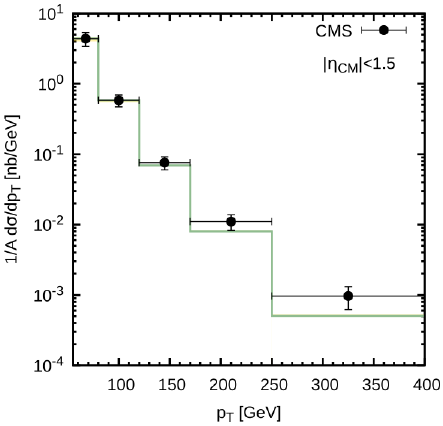}
\includegraphics[width=7.9cm]{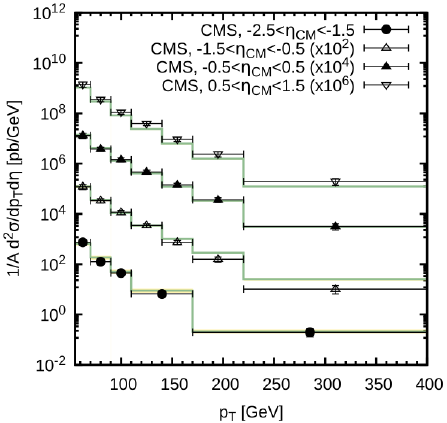}
\caption{Differential cross sections of $c$ (left) and $b$ (right) jets production as functions of transverse momenta of the leading jet measured at $\sqrt s_\text{NN}=5.02$~TeV. The uncertainty band corresponds to estimates made as described in the text. Experimental data are from~\cite{CMS-cjets,CMS-bjets}.
}
\label{fig-jets}
\end{center}
\end{figure}

We start from charm and beauty jet production at proton-lead collisions at $\sqrt s_\text{NN}=5.02$~TeV. 
The CMS data on $c$-jet transverse momentum spectra\cite{CMS-cjets} refer to the pseudorapidity 
region (in nucleon-nucleon center-of-mass\footnote{The laboratory frame is defined with a 
shift of $\eta_\text{lab}=\eta_\text{CM}+0.465$ with the positive direction corresponding to the proton beam direction.} frame)
of $|\eta_\text{CM}|<1.5$. The data on $b$-jet production\cite{CMS-bjets} were 
measured at four different pseudorapidity differences: $-2.5<|\eta_\text{CM}|<-1.5$, $-1.5<|\eta_\text{CM}|<-0.5$, $-0.5<|\eta_\text{CM}|<0.5$ and $0.5<|\eta_\text{CM}|<1.5$. 
Both these dataset correspond to jet transverse momentum $55<p_T<400$~GeV. 
Our predictions %for heavy jet transverse momentum distributions 
are shown in Fig.~\ref{fig-jets}
in comparison with the CMS data\cite{CMS-cjets, CMS-bjets}.
The shaded bands (which, in fact, are quite narrow) represent our theoretical uncertainties estimated as discussed above. 
One can see that overall description of the data is reasonable good in all the pseudorapidity regions.
The shape and absolute normalization of both measured charm and beauty
jets are reproduced well, except, may be, last $p_T$ bin ($p_T \sim 300$~GeV), where effects of 
parton showers and/or hadronization can play a role\footnote{Studying of these effects is out of range of the present paper.}.

\begin{figure}
\begin{center}
\includegraphics[width=7.7cm]{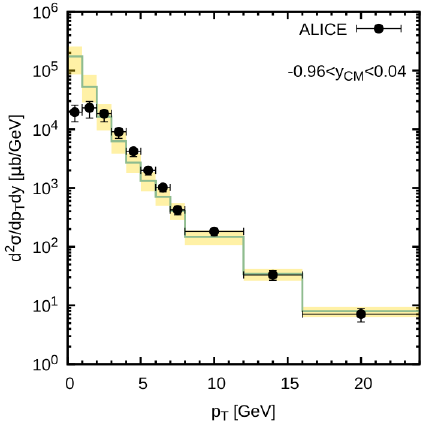}
\includegraphics[width=7.9cm]{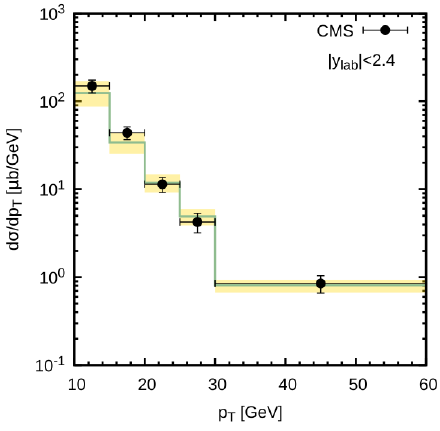}
\includegraphics[width=7.9cm]{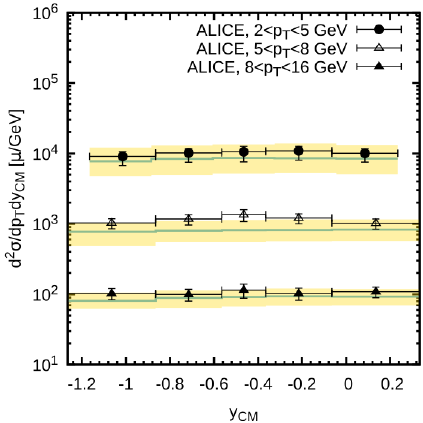}
\includegraphics[width=7.9cm]{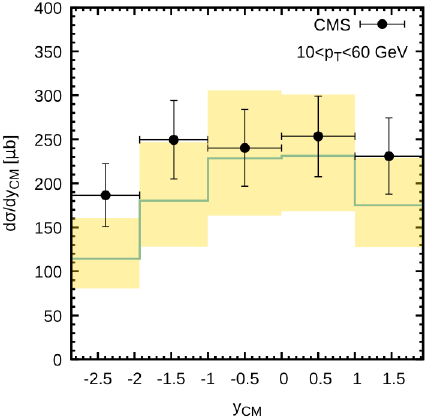}
\caption{Differential cross sections of $D^0$ (left) and $B^+$ (right) jets production as functions of transverse momenta of the meson (upper panels) and their rapidity (lower panels) measured at $\sqrt s_\text{NN}=5.02$~TeV. The uncertainty band corresponds to estimates made as described in the text. Experimental data are from~\cite{ALICE-D,CMS-B}.
}
\label{fig-mesons}
\end{center}
\end{figure}

To test lower transverse momenta we turn to data on $D$ and $B$ meson production in proton-lead collisions at the same energy. 
So, ALICE collaboration provided us with $D^0$ production data\cite{ALICE-D} measured in the kinematical 
region $0<p_T<24$ and $-0.96<y_\text{CM}<0.04$. The $D^0$ rapidity spectra 
were measured at three different regions of transverse momenta: $2<p_T<5$, $5<p_T<8$ and $8<p_T<16$~GeV, where 
rapidity region was extended then to $-1.265<y_\text{CM}<0.335$. 
The CMS data on $B^+$ meson production\cite{CMS-B} were taken at $10<p_T<60$~GeV and $|y_\text{lab}|<2.4$. 
To convert $c$ or $b$ quarks to $D^0$ or $B^+$ mesons we employ the standard Peterson fragmentation function
with corresponding shape parameters $\epsilon_c=0.06$ and $\epsilon_b=0.006$, 
which are often used in the NLO pQCD calculations.
Following to\cite{PDG}, we set branching ratios $B(c\to D^0)=0.559$ and $B(b\to B^+)=0.408$.
The results of performed calculations are shown in Fig.~\ref{fig-mesons}. 
One can see that our predictions for all observables are consistent with the data within the 
theoretical and experimental uncertainties.
There is only disagreement with the ALICE data on $D^0$ transverse momentum distribution at very low $p_T \leq m(D^0)$.
However, we would like to point out that in order to obtain the reliable predictions at low $p_T$
the large logarithmic terms $\sim \alpha_s^n\ln^n m_Q/p_T$ have to be resummed.
It can be done using, for example, special soft gluon resummation technique\cite{SoftGluonResummation-I, SoftGluonResummation-II}
or Collins-Soper-Sterman framework\cite{CSS-1,CSS-2,CSS-3} (see also\cite{TMD-Review} for more information).
This issue is not considered here.
We note also that problematic low $p_T$ region, $p_T \leq m(D^0)$, is excluded from $D^0$ rapidity measurements, see Fig.~\ref{fig-mesons} (lower panels).

Another important observable is the nuclear modification factor $R^Q_{pA}$, describing the nuclear medium influence on the production dynamics:
\begin{equation}
R_{pA}=\frac{1}{A} {\sigma(pA \to Q + X) \over \sigma(pp \to Q + X)},
\end{equation}
where $A = 208$ is the number of nucleons in the Pb nucleus. %In our calculations it is purely defined by the TMD difference.
Our results for $R^Q_{pA}$ (with $Q = c$ or $b$) 
are summarized in Table~1. All the estimated values are close to unity and our predictions
agree with the experimental results within uncertainties.

%, except of $D^0$ case, where low-$p_T$ behaviour seems to be crucial for the description of the data. 
%Our approach is barely applicable to very low $p_T$, so it fails to describe this observable.

\begin{table} \footnotesize 
\label{table1}
\begin{center}
\begin{tabular}{|c|c|c|}
\hline
 & & \\
Process   & experiment & theory \\
 & & \\
\hline 
 & & \\
$c$-jet production   & $0.92\pm 0.13$\cite{CMS-cjets} & $0.77^{+0.11}_{-0.07}$ \\
 & & \\
\hline 
 & & \\
$b$-jet production   & $1.22\pm 0.31$\cite{CMS-bjets} & $0.82^{+0.14}_{-0.04}$ \\
 & & \\
\hline 
 & & \\
%$D^0$ production~\cite{ALICE-D}   & $0.89^{+0.17}_{-0.21}$ & $0.45^{+0.09}_{-0.02}$ \\\hline 
$B^+$ production   & $1.11^{+0.38}_{-0.35}$\cite{CMS-B} & $0.77^{+0.02}_{-0.03}$ \\
 & & \\
\hline 
\end{tabular}
\end{center}
\caption{Calculated nuclear modification factors $R^Q_{pA}$ %in the fiducial regions %calculated in the present work with nKLSZ and KLSZ TMDs 
in comparison with the LHC data. The estimated theoretical uncertainties come from the scale variation as it is described in the text.}
\end{table}

Thus, our calculations show that the derived expressions for TMD gluon densities
in a proton and nuclei provide a reasonably good description of LHC data 
for heavy flavor production in proton-lead collisions at the LHC,
which is extremely sensitive to the gluon content of colliding particles.
It is important for future investigations of proton-nucleus and 
nucleus-nucleus interactions in the TMD-based framework.
Note that the consideration could be improved
by taking into account Gross-Llewellyn-Smith and Gottfried sum rules
for valence and nonsinglet quark distributions and 
momentum conservation for singlet quark and gluon densities
similar to that as it was already done\cite{PDFsSumRules}.
We plan to perform such derivation in a forthcoming study.
Also we plan to incorporate the calculated nTMDs into the Monte-Carlo event generator \textsc{pegasus}\cite{PEGASUS}
that significantly extends its feasibilities. 

\section{Conclusions} \noindent

We have derived analytical expressions for the TMD gluon and sea quark
densities in nuclei at leading order of QCD running coupling.
The calculations are performed in the framework of the rescaling model and
Kimber-Martin-Ryskin prescription, where the Bessel-inspired behavior of parton densities at 
small Bjorken $x$ values, obtained in the case of flat initial conditions in the
double scaling QCD approximation, is applied. 
Then we apply the obtained expressions to inclusive heavy flavor production in proton-lead collisions at the LHC,
which is known to be sensitive to the gluon content of colliding particles.
The calculations were performed in the framework of the $k_T$-factorization QCD approach and
based on the dominating off-shell gluon-gluon fusion $g^* g^* \to Q\bar Q$ subprocess.
The analysis covers the transverse momentum and rapidity spectra of charm and beauty jets as well as $D^0$ and $B^+$ mesons.
We find a good agreement of our results with 
latest experimental data collected by the CMS and ALICE Collaborations at $\sqrt s = 5.02$~GeV
with the theoretical and experimental uncertainties.
It is important for future studies of proton-nucleus and nucleus-nucleus  interactions
within the TMD-based approaches.

\section*{Acknowledgements} \indent

We thank S.P.~Baranov for his interest, useful discussions
and important remarks. 
Our study was in part supported by the 
Russian Science Foundation under grant~22-22-00387.
The work of A.V.K. and A.V.L. was supported in part by
the CAS President’s International Fellowship Initiative. % (Grants No. 2017VMA0040 and No. 2017VMA0040, respectively). They thank Institute of Modern Physics for invitation.}
  
\bibliography{bjet-nucl}

\begin{thebibliography}{10}

\bibitem{EPPS22}
K.J.~Eskola{,} P.~Paakkinen{,} H.~Paukkunen{,} C.A.~Salgado{,} Eur. Phys. J. C
  {\bf 82}{,}~413 (2022).

\bibitem{nNNPDF3}
R.A.~Khalek{,} R.~Gauld{,} T.~Giani{,} E.R.~Nocera{,} T.R.~Rabemananjara{,}
  J.~Rojo{,} Eur. Phys. J.C {\bf 82}{,}~507 (2022).

\bibitem{nCTEQ15}
K.~Kovarik{,} A.~Kusina{,} T.~Jezo{,} D.B.~Clark{,} C.~Keppel{,} F.~Lyonnet{,}
  J.G.~Morfin{,} F.I.~Olness{,} J.F.~Owens{,} I.~Schienbein{,} J.Y.~Yu{,} Phys.
  Rev. D {\bf 93}{,}~085037 (2016).

\bibitem{nCTEQ15upd}
P.~Duwent\"aster{,} T.~Jezo{,} M.~Klasen{,} K.~Kovarik{,} A.~Kusina{,}
  K.F.~Muzakka{,} F.I.~Olness{,} R.~Ruiz{,} I.~Schienbein{,} J.Y.~Yu{,} Phys.
  Rev. D {\bf 105}~114043{,} (2022).

\bibitem{nFactorization-1}
J.-W.~Qiu{,} I.~Vitev{,} Phys. Lett. B {\bf 632}{,}~507 (2006).

\bibitem{nFactorization-2}
H.~Fujii{,} K.~Watanabe{,} Nucl.Phys. A {\bf 920}{,}~78 (2013).

\bibitem{nFactorization-3}
Z.-B.~Kang{,} I.~Vitev{,} E.~Wang{,} H.~Xing{,} C.~Zhang{,} Phys. Lett. B {\bf
  740}{,}~23 (2015).

\bibitem{nFactorization-4}
B.~Guiot{,} B.Z.~Kopeliovich{,} Phys. Rev. C {\bf 102}{,}~045201 (2020).

\bibitem{Nuclear-Review}
A.~Andronic{,} F.~Arleo{,} R.~Arnaldi{,} A.~Beraudo{,} E.~Bruna{,}
  D.~Caffarri{,} Z.~Conesa del Valle{,} J.G.~Contreras{,} T.~Dahms{,}
  A.~Dainese{,} M.~Djordjevic{,} E.G.~Ferreiro{,} H.~Fujii{,} P.B.~Gossiaux{,}
  R.~Granier de Cassagnac{,} C.~Hadjidakis{,} M.~He{,} H.~van Hees{,}
  W.A.~Horowitz{,} R.~Kolevatov{,} B.Z.~Kopeliovich{,} J.P.~Lansberg{,}
  M.P.~Lombardo{,} C.~Lourenco{,} G.~Martinez-Garcia{,} L.~Massacrier{,}
  C.~Mironov{,} A.~Mischke{,} M.~Nahrgang{,} M.~Nguyen{,} J.~Nystrand{,}
  S.~Peigne{,} S.~Porteboeuf-Houssais{,} I.K.~Potashnikova{,}
  A.~Rakotozafindrabe{,} R.~Rapp{,} P.~Robbe{,} M.~Rosati{,} P.~Rosnet{,}
  H.~Satz{,} R.~Schicker{,} I.~Schienbein{,} I.~Schmidt{,} E.~Scomparin{,}
  R.~Sharma{,} J.~Stachel{,} D.~Stocco{,} M.~Strickland{,} R.~Tieulent{,}
  B.A.~Trzeciak{,} J.~Uphoff{,} I.~Vitev{,} R.~Vogt{,} K.~Watanabe{,}
  H.~Woehri{,} P.~Zhuang{,} Eur. Phys. J. C {\bf 76}{,} 107~(2016).

\bibitem{EMC}
EMC Collaboration{,} Phys. Lett. B {\bf 123}{,}~275 (1983).

\bibitem{EMC-Review-1}
N.N.~Nikolaev{,} Sov. Phys. Usp. {\bf 24}{,} 531 (1981);\\ V.~Barone {\it et
  al.}{,} Z.~Phys. C {\bf 58}{,} 541 (1993);\\ N.N.~Nikolaev{,}
  B.G.~Zakharov{,} Z. Phys. C {\bf 49}{,} 607~(1991).

\bibitem{EMC-Review-2}
M.~Arneodo{,} Phys. Rept. {\bf 240}{,} 301 (1994);\\ P.R.~Norton{,} Rept. Prog.
  Phys. {\bf 66}{,}~1253 (2003).

\bibitem{DGLAP}
V.N. Gribov{,} L.N.~Lipatov{,} Sov. J. Nucl. Phys. {\bf 15}{,} 438 (1972); \\
  L.N.~Lipatov{,} Sov. J. Nucl. Phys. {\bf 20}{,} 94 (1975); \\ G.~Altarelli{,}
  G.~Parisi{,} Nucl. Phys. B {\bf 126}{,} 298 (1977); \\ Yu.L. Dokshitzer{,}
  Sov. Phys. JETP {\bf 46}{,}~641 (1977).

\bibitem{Kulagin}
S.A.~Kulagin{,} Phys. Part. Nucl. {\bf 50}{,} 506 (2019);\\ S.A.~Kulagin{,} EPJ
  Web~Conf. \textbf{138}{,} 01006~(2017).

\bibitem{RescalingModel}
R.L.~Jaffe {\it et al.,}~Phys. Lett. B {\bf 134}{,} 449 (1984);\\
  O.~Nachtmann{,} H.J.~Pirner{,} Z. Phys. C {\bf 21}{,} 277 (1984);\\
  F.E.~Close~{\it et al.,} Phys. Rev. D {\bf 31} 1004{,}~(1985).

\bibitem{RescalingModelAssumptions}
F.E.~Close{,} R.G.~Roberts{,} G.G.~Ross{,} Phys. Lett. B {\bf 129}{,} 346
  (1983);\\ R.L.~Jaffe{,} Phys. Rev. Lett. {\bf 50}{,}~228 (1983).

\bibitem{RescalingModel-Review}
R.L.~Jaffe{,} arXiv{:}2212.05616 [hep{-}ph].

\bibitem{RescalingModel-Extension-1}
A.V.~Kotikov{,} B.G.~Shaikhatdenov{,} P.~Zhang{,} Phys. Rev. D {\bf
  96}{,}~114002 (2017).

\bibitem{RescalingModel-Extension-2}
A.V.~Kotikov{,} B.G.~Shaikhatdenov{,} P.~Zhang{,} Phys. Part. Nucl. Lett. {\bf
  16}{,} 311 (2019)~[1811.05615 [hep ph]].

\bibitem{RescalingModel-Extension-3}
N.A.~Abdulov{,} A.V.~Kotikov{,} A.V.~Lipatov{,} Phys. Part. Nucl.~Lett.
  \textbf{20}{,} 557 (2023);\\ A.V.~Kotikov{,} B.G.~Shaikhatdenov{,}
  P.~Zhang{,} EPJ Web Conf.~\textbf{204}{,} 05002~(2019).

\bibitem{RescalingModel-LinearlyPolarizedGluons}
N.A.~Abdulov{,} X.~Chen{,} A.V.~Kotikov{,} A.V.~Lipatov{,} arXiv{:}2310.10496
  [hep{-}ph].

\bibitem{LinearlyPolarizedGluons}
N.A.~Abdulov{,} X.~Chen{,} A.V.~Kotikov{,} A.V.~Lipatov{,} arXiv{:}2310.08107
  [hep{-}ph].

\bibitem{KMR-LO}
M.A.~Kimber{,} A.D.~Martin{,} M.G.~Ryskin{,} Phys. Rev. D {\bf 63}{,} 114027
  (2001);\\ G.~Watt{,} A.D.~Martin{,} M.G.~Ryskin{,} Eur. Phys. J. C {\bf
  31}{,}~73 (2003).

\bibitem{KMR-NLO}
A.D.~Martin{,} M.G.~Ryskin{,} G.~Watt{,} Eur. Phys. J. C {\bf 66}{,}~163
  (2010).

\bibitem{TMD-Review}
R.~Angeles-Martinez{,} A.~Bacchetta{,} I.I.~Balitsky{,} D.~Boer{,}
  M.~Boglione{,} R.~Boussarie{,} F.A.~Ceccopieri{,} I.O.~Cherednikov{,}
  P.~Connor{,} M.G.~Echevarria{,} G.~Ferrera{,} J.~Grados Luyando{,}
  F.~Hautmann{,} H.~Jung{,} T.~Kasemets{,} K.~Kutak{,} J.P.~Lansberg{,}
  A.~Lelek{,} G.I.~Lykasov{,} J.D.~Madrigal Martinez{,} P.J.~Mulders{,}
  E.R.~Nocera{,} E.~Petreska{,} C.~Pisano{,} R.~Placakyte{,} V.~Radescu{,}
  M.~Radici{,} G.~Schnell{,} I.~Scimemi{,} A.~Signori{,} L.~Szymanowski{,}
  S.~Taheri Monfared{,}~F.F. van~der~Veken{,} H.J.~van~Haevermaet{,} P.~van
  Mechelen{,} A.A.~Vladimirov{,} S.~Wallon{,} Acta Phys. Polon. B {\bf 46}{,}
  2501~(2015).

\bibitem{TMD-map}
S.P.~Baranov{,} H.~Jung{,} A.V.~Lipatov{,} M.A.~Malyshev{,} Eur. Phys. J. C
  {\bf 77}{,}~2 (2017).

\bibitem{kt-factorization}
L.V.~Gribov{,} E.M.~Levin{,} M.G.~Ryskin{,} Phys. Rep. {\bf 100}{,} 1 (1983){;}
  \\ E.M.~Levin{,} M.G.~Ryskin{,} Yu.M.~Shabelsky{,} A.G.~Shuvaev{,} Sov. J.
  Nucl. Phys. {\bf 53}{,}~657 (1991).

\bibitem{HighEnergyFactorization}
S.~Catani{,} M.~Ciafaloni{,} F.~Hautmann{,} Nucl. Phys. B {\bf 366}{,} 135
  (1991){;} \\ J.C.~Collins{,} R.K.~Ellis{,} Nucl. Phys. B {\bf 360}{,}~3
  (1991).

\bibitem{CMS-B}
CMS Collaboration{,} Phys. Rev. Lett. {\bf 116}{,}~032301 (2016).

\bibitem{CMS-bjets}
CMS Collaboration{,} Phys. Lett. B {\bf 754}{,}~59 (2016).

\bibitem{CMS-cjets}
CMS Collaboration{,} Phys. Lett. B {\bf 772}{,}~306 (2017).

\bibitem{ALICE-D}
ALICE Collaboration{,} Phys. Rev. Lett. {\bf 113}{,}~232301 (2014).

\bibitem{TMD-DAS-1}
A.V.~Kotikov{,} A.V.~Lipatov{,} B.G.~Shaikhatdenov{,} P.~Zhang{,} JHEP {\bf
  02}{,}~028 (2020).

\bibitem{TMD-DAS-2}
A.V.~Kotikov{,} A.V.~Lipatov{,} P.~Zhang{,} Phys. Rev. D {\bf 104}{,}~054042
  (2021).

\bibitem{DAS-Theory-1}
R.D.~Ball{,} S.~Forte{,} Phys. Lett. B {\bf 336}{,}~77 (1994).

\bibitem{DAS-Theory-2}
A.V.~Kotikov{,} G.~Parente{,} Nucl. Phys. B {\bf 549}{,}~242 (1999).

\bibitem{DAS-Theory-3}
A.Yu.~Illarionov{,} A.V.~Kotikov{,} G.~Parente{,} Phys. Part. Nucl. {\bf
  39}{,}~307 (2008).

\bibitem{DAS-Theory-4}
L.~Mankiewicz{,} A.~Saalfeld{,} T.~Weigl{,} Phys. Lett. B {\bf 393}{,}~175
  (1997).

\bibitem{PDFs-DAS-1}
G.~Cvetic{,} A.Yu.~Illarionov{,} B.A.~Kniehl{,} A.V.~Kotikov{,} Phys. Lett. B
  {\bf 679}{,}~350 (2009).

\bibitem{frozen}
B.~Badelek{,} J.~Kwiecinski{,} A.~Stasto{,} Z. Phys. C {\bf 74}{,} 297
  (1997);\\ A.V.~Kotikov{,} A.V.~Lipatov{,} N.P.~Zotov{,} J.Exp.Theor.Phys.
  {\bf 101}{,}~811 (2005).

\bibitem{TMDLib2}
N.A.~Abdulov{,} A.~Bacchetta{,} S.P.~Baranov{,} A.~Bermudez Martinez{,}
  V.~Bertone{,} C.~Bissolotti{,} V.~Candelise{,} L.I.~Estevez Banos{,}
  M.~Bury{,} P.L.S.~Connor{,} L.~Favart{,} F.~Guzman{,} F.~Hautmann{,}
  M.~Hentschinski{,} H.~Jung{,} L.~Keersmaekers{,} A.V.~Kotikov{,} A.~Kusina{,}
  K.~Kutak{,} A.~Lelek{,} J.~Lidrych{,} A.V.~Lipatov{,} G.I.~Lykasov{,}
  M.A.~Malyshev{,} M.~Mendizabal{,} S.~Prestel{,} S.~Sadeghi Barzani{,}
  S.~Sapeta{,} M.~Schmitz{,} A.~Signori{,} G.~Sorrentino{,} S.~Taheri
  Monfared{,}~A. van~Hameren{,} A.M.~van Kampen{,} M.~Vanden Bemden{,}
  A.~Vladimirov{,} Q.~Wang{,} H.~Yang{,} Eur. Phys. J. C {\bf 81}{,}
  752~(2021).

\bibitem{HF-ZLS}
N.P.~Zotov{,} A.V.~Lipatov{,} V.A.~Saleev{,} Phys. Atom. Nucl. {\bf 66}{,}~755
  (2003).

\bibitem{HF-JKLZ}
H.~Jung{,} M.~Kraemer{,} A.V.~Lipatov{,} N.P.~Zotov{,} JHEP {\bf 01}{,} 085
  (2011){;} \\ H.~Jung{,} M.~Kraemer{,} A.V.~Lipatov{,} N.P.~Zotov{,} Phys.
  Rev. D {\bf 85}{,}~034035 (2012).

\bibitem{PDFsSumRules}
N.A.~Abdulov{,} A.V.~Kotikov{,} A.V.~Lipatov{,} Particles {\bf 5}{,}~535
  (2022).

\bibitem{PDG}
PDG Collaboration{,} Prog. Theor. Exp. Phys. 2022{,}~083C01 (2022).

\bibitem{SoftGluonResummation-I}
Yu.L.~Dokshitzer{,} D.~D'yakonov{,} S.I.~Troyan{,} Phys. Lett. B {\bf
  79}{,}~269 (1978).

\bibitem{SoftGluonResummation-II}
G.~Parisi{,} R.~Petronzio{,} Nucl. Phys. B {\bf 154}{,}~427 (1979).

\bibitem{CSS-1}
J.C.~Collins{,} D.E.~Soper{,} G.F.~Sterman{,} Nucl. Phys.~B \textbf{250}{,} 199
  (1985){;}\\ J.C.~Collins{,} D.E.~Soper{,} Nucl. Phys. B~\textbf{223}{,}
  381~(1983).

\bibitem{CSS-2}
J.C.~Collins{,} D.E.~Soper{,} Nucl. Phys.~B \textbf{194}{,} 445 (1982){;}\\
  J.C.~Collins{,} D.E.~Soper{,} Nucl. Phys. B~\textbf{197}{,} 446~(1982).

\bibitem{CSS-3}
R.D.~Ball \textit{et al.}{,} JHEP~\textbf{04}{,} 040~(2015).

\bibitem{PEGASUS}
A.V.~Lipatov{,} M.A.~Malyshev{,} S.P.~Baranov{,} Eur. Phys. J. C {\bf
  80}{,}~330 (2020).

\end{thebibliography}

\end{document}